\documentstyle[aps,pra,multicol,epsf]{revtex}
\begin{document}
\draft

\title{A Scheme to Probe the Decoherence of a Macroscopic Object}
\author{S.\ Bose, K.\ Jacobs and P.\ L.\ Knight}
\address{Optics Section, The Blackett Laboratory, Imperial College, London SW7 2BZ, England}

\maketitle
\begin{abstract}
We propose a quantum optical version of Schr\"{o}dinger's famous gedanken experiment in which the state of a microscopic system (a cavity field) becomes entangled with and disentangled from the state of a massive object (a movable mirror). Despite the fact that a mixture of Schr\"{o}dinger cat states is 
produced during the evolution (due to the fact that the macroscopic mirror starts off in a thermal state), this setup allows us to systematically probe the rules by which a superposition of spatially separated states of a macroscopic object decoheres. The parameter regime required to test  environment-induced decoherence models is found to be close to those currently realizable, while that required to detect gravitationally induced collapse is well beyond current technology.

\end{abstract}

\begin{multicols}{2}
\section{Introduction}
 Quantum mechanical superpositions of macroscopically 
distinguishable states of a macroscopic object decay rapidly due to the strong coupling of the object with its
environment. This process is called environment-induced decoherence (EID) \cite{zkpt}. There are many assumptions involved in 
modelling the EID of a macroscopic object. For example, the nature of the coupling of a macroscopic object to
its environment is generally taken to be a linear \cite{zk} or a nonlinear \cite{hu} function of the position operator of the object. Assumptions are also made about the environment. Based on these assumptions, various explicit formulae have been derived for the dependence of the decoherence time scale on various parameters of the system, its environment, and the spatial separation between the superposed components~\cite{zkpt,zk,zk1,hu}. Obviously, the most appropriate model can be selected only through experimentation. Decoherence formulae relevant to the quantum optical domain~\cite{mil} are
now beginning to be tested experimentally \cite{Har}. As far as quantum objects bearing mass are concerned, decoherence has been investigated for the motional states of ions in a trap \cite{mon}. There have also
been other interesting suggestions for testing decoherence experimentally \cite{leg,Hm,angl}. However, as yet no one has managed to test the 
the rules of decoherence of a superposition of spatially separated states
of a macroscopic object. This is, presumably, because of the
implicit assumption that one actually needs to prepare a superposition of distinct
states of a macroscopic object in order to probe the
rules of its decoherence. Such a superposition is
extremely difficult to prepare because of the
difficulty of obtaining a macroscopic system in a
pure quantum state. In this article, we propose a scheme that 
will allow us to probe the rules of decoherence of
a superposition of states
of a macroscopic object without
actually creating such a superposition. We also show that it
requires experimental parameters that are close to the potentially realizable domain.

 Besides EID, there also exist a set of collapse models~\cite{grw,pen} developed to resolve the measurement problem of quantum mechanics. Whether
such a mechanism is really necessary or whether some reformulation of quantum mechanics such as the histories approach~\cite{his} suffices, is an open question~\cite{debates}. Some experiments to detect such mechanisms have been suggested~\cite{Hm,bose}, and some preexisting experimental data have been analysed ~\cite{stnz,perl}. In particular, atomic interferometry experiments provide
a great potential to put bounds on such models \cite{stnz}. However, there is no direct evidence for their existence. We calculate the parameter regime required if our experiment is to probe such models, and show that this is a much more difficult task than probing EID. 

      Our experiment is based on applying the ideas used by Schr\"{o}dinger
in his famous gedanken experiment involving a cat \cite{sr} to a certain
quantum optical system. Obvious differences arise as our set up is meant
to be realistic. We begin by recapitulating Schr\"{o}dinger's technique
and describing qualitatively what happens when such a technique is
applied to our set up.

\section{Schr\"{o}dinger's method for creating macroscopic superpositions}
The basic idea used by Schr\"{o}dinger to create macroscopic superpositions
was to entangle the states of a microscopic and a macroscopic system \cite{sr}. It is
easy to put the state of a microscopic system (which follows quantum mechanics beyond any controversy) into a superposition of distinct states.
Subsequently, this system is allowed to interact with a macroscopic system
to propel it to macroscopically distinct states corresponding to the different
superposed states of the microscopic system. In Schr\"{o}dinger's case, the
microscopic system was a radioactive atom, while the macroscopic system
was a cat. In this paper we propose to apply exactly the Schr\"{o}dinger technique to
a cavity field (a microscopic system) coupled to a movable mirror (a macroscopic
system). However, there are differences of such a realistic
version of Schr\"{o}dinger's thought experiment from his
original version. We will enumerate these
problems below and outline why our experiment
can still achieve its target (testing the decoherence of
superpositions of states of a macroscopic object) despite being
quite different from  Schr\"{o}dinger's original experiment.             

      Firstly, as yet no technique exists to put a macroscopic oscillator in a
pure coherent state (though some progress has been made in cooling of such objects \cite{tom}). Thus, unlike the cat in Schr\"{o}dinger's original experiment, the macroscopic mirror cannot start off in a pure state, and in
general, will start off in a thermal state. So only a
mixture of Schr\"{o}dinger's cat states can ever be created through unitary
evolutions. So the primary aim of Schr\"{o}dinger's experiment (creating macroscopic superpositions) cannot be achieved this way. However, our aim is to test the decoherence of states of a macroscopic object and not to create a pure Schr\"{o}dinger's cat state. An important point to realize is that the
former {\em can} be done {\em without} necessarily doing the latter. We
shall show that the
state of the cavity field (the system on which we actually perform our measurements) at the end of our experiment is
same {\em irrespective} of whether the macroscopic object (the movable mirror) coupled to it starts off in a thermal state or a
coherent state. This is due to the specific nature of the coupling between the field
and the mirror. Thus the mixture of Schr\"{o}dinger's cat states produced
has the same observational consequences as a pure Schr\"{o}dinger's cat as
far as our scheme is concerned.

     Secondly, it appears that the decoherence of a superposition of states of a macroscopic mirror (which can never be made as isolated
as the cat in Schr\"{o}dinger's thought experiment) 
will be too fast to detect. To circumvent 
this, one must note that the decoherence rate of a certain superposition of states of an object increases with the 
spatial separation between these states \cite{zkpt,zk,zk1}. So the amount by which the decoherence rate
increases due to the macroscopic nature (large mass) of the object can
always be offset by reducing the spatial separation between the superposed states.

    Thirdly, in Schr\"{o}dinger's case, the coupling between radioactivity and
the cat involved highly nonlinear biological processes. So even a small trigger
of radioactive decay was sufficient (via the breaking of the poison vial)
to produce as much of a change in a cat as killing it. Can we get such a
nonlinear coupling to produce a drastic change in the state of the
macroscopic mirror from small changes in the state of the cavity field? The answer to this is that it is really not necessary to have a drastic change in the
state of the macroscopic mirror. Even a superposition of macroscopically
non-discernible states
is sufficient to produce a detectable rate of decoherence if the
the mirror is sufficiently macroscopic.

\section{The system under consideration}   

We consider a micro-cavity with one fixed and one movable mirror. The cavity contains  a single mode of an electromagnetic field (frequency $\omega_0$ and annihilation 
operator denoted by $a$) that couples to the movable mirror (which is treated as a quantum harmonic oscillator of frequency $\omega_m$ 
and annihilation operator denoted by $b$). This system has already been studied quite extensively \cite{Tombesi,us,Kurt} and relevant Hamiltonian~\cite{Tombesi} is

\begin{equation}
\label{g}
  H = \hbar\omega_0~a^\dagger a ~+~\hbar\omega_m~b^\dagger 
b~-~\hbar g~a^\dagger a (b+b^\dagger)
\end{equation}

where
\begin{equation}
\label{gpar}
  g~=~\frac{\omega_0}{L} ~\sqrt\frac{\hbar}{2m\omega_m} ~,
\end{equation} 
and  $L$ and $m$ are the length of the cavity and mass of 
the movable mirror respectively. For the moment we consider the system to be
totally isolated.
If the field inside the cavity was initially in a number state $|n\rangle_{\mbox{\scriptsize c}}$ and the mirror was initially in a coherent state $|\beta \rangle_{\mbox{\scriptsize m}}$, then
at any later time $t$ the mirror will be in the coherent state \cite{us}
\begin{equation}
\label{phi1}
  |\phi_n (t)\rangle_{\mbox{\scriptsize m}} =|~\beta e^{-i\omega_m t} + \kappa n(1-e^{-i\omega_m t})~\rangle_{\mbox{\scriptsize m}} ~,
\end{equation}
where $\kappa = g/\omega_m$.
Thus, in effect the mirror undergoes an oscillation with a frequency $\omega_m$
and an amplitude dependent on the Fock state inside the cavity. This feature of the mirror dynamics
gives us the basic idea of the paper. A superposition of
two different Fock states is created inside the cavity so that the mirror is driven to an oscillation of different amplitude corresponding to each of these Fock states. As the mirror is a macroscopic object, this situation can be
regarded as a version of Schr\"{o}dinger's cat experiment. Of course,
in practice, only a mixture of several Schr\"{o}dinger's cats is created
because the mirror starts off in a thermal state instead of starting in a
single coherent state $|\beta \rangle_{\mbox{\scriptsize m}}$.

\section{The proposed scheme}

     We propose to start with the cavity field prepared in the
initial superposition of Fock states
\begin{equation}
\label{ini1}  
     |\psi(0) \rangle_{\mbox{\scriptsize c}} = \frac{1}{\sqrt{2}}(|0\rangle_{\mbox{\scriptsize c}} - |n\rangle_{\mbox{\scriptsize c}})  .
\end{equation}
Methods of preparing the cavity field in such states has been described in Refs.\cite{par,vog,lw}. When discussing experimental parameters we will choose $n=1$, which is the simplest to prepare. The initial state of the movable mirror will be 
taken to be a thermal
state at some temperature $T$, and is given by the density operator  
\begin{equation}
\rho_{\mbox{\scriptsize m}} = \frac{1}{\pi \bar{n}} \int (|\beta\rangle \langle \beta|)_{\mbox{\scriptsize m}} \exp(-|\beta|^2/\bar{n}) ~ d^2\beta ,
\label{th}   
\end{equation}
where 
\begin{equation}
   \bar{n}= \frac{1}{e^{\hbar \omega_m / k_{\mbox{\tiny B}} T} - 1}  ,
\end{equation}
and $|\beta \rangle_m$ represents a coherent state of the mirror corresponding
to amplitude $\beta$ and $k_{\mbox{\tiny B}}$ is the Boltzmann constant.
Eq.(\ref{phi1}) and the initial states given by
Eqs.(\ref{ini1}) and (\ref{th}) imply that at any time $t$, in the absence of any external
environment, the joint density operator describing the cavity mode and the
mirror is given by
\begin{eqnarray}
\label{cat}
\rho(t)_{\mbox{\scriptsize {c+m}}} &=& \frac{1}{2 \pi \bar{n}} \int (\rho(t)_{00}-\rho(t)_{0n} \nonumber \\
&-& \rho(t)_{n0}+\rho(t)_{nn})_{\mbox{\scriptsize {c+m}}} \exp(-|\beta|^2/\bar{n}) ~ d^2\beta ,   
\end{eqnarray}
where
\begin{mathletters}
\begin{eqnarray}
\rho(t)_{00} & = & (|0\rangle \langle 0|)_{\mbox{\scriptsize c}} \otimes (|\phi_0 (t)\rangle \langle \phi_0(t)|)_{\mbox{\scriptsize m}} \label{e1}, \\
\rho(t)_{0n} & = & (|0\rangle \langle n|)_{\mbox{\scriptsize c}} \otimes (|\phi_0 (t)\rangle \langle \phi_n(t)|)_{\mbox{\scriptsize m}} \nonumber \\
             &  & \times e^{-i \kappa^2 n^2 (\omega_mt-\sin{\omega_mt})} \label{e2} , \\
\rho(t)_{n0} & = & (|n\rangle \langle 0|)_{\mbox{\scriptsize c}} \otimes (|\phi_n (t)\rangle \langle \phi_0(t)|)_{\mbox{\scriptsize m}} \nonumber \\ 
            &  & \times e^{i \kappa^2 n^2 (\omega_mt-\sin{\omega_mt})} , \label{e3} \\
\rho(t)_{nn} & = &  (|n\rangle \langle n|)_{\mbox{\scriptsize c}} \otimes (|\phi_n (t)\rangle \langle \phi_n(t)|)_{\mbox{\scriptsize m}} \label{e4},
\end{eqnarray}
\end{mathletters}
The phase factors $ e^{\pm i \kappa^2 n^2 (\omega_mt-\sin{\omega_mt})} $ in Eqs.(\ref{e2}) and (\ref{e3}) derive from a Kerr like term in 
the time evolution operator corresponding to the Hamiltonian $H$, which has been
evaluated in Refs.\cite{Tombesi,us}. Note that there
are absolutely no assumptions involved while writing Eq.(\ref{cat}).
However, the coherent state basis expansion of the initial
thermal state of the mirror (Eq.(\ref{th})) has been used for a specific
purpose. The terms $\rho(t)_{00}$, $\rho(t)_{0n}$, $\rho(t)_{n0}$ and
$\rho(t)_{nn}$ appear in Eq.(\ref{cat}) only if such an expansion is
made. The effect of decoherence on such terms is already well
studied \cite{zk,zk1} and the specific form of Eq.(\ref{th})
allows us to simply utilize these known results.
     
\begin{figure} 
\begin{center} 
\leavevmode 
\epsfxsize=1.5cm 
\epsfbox{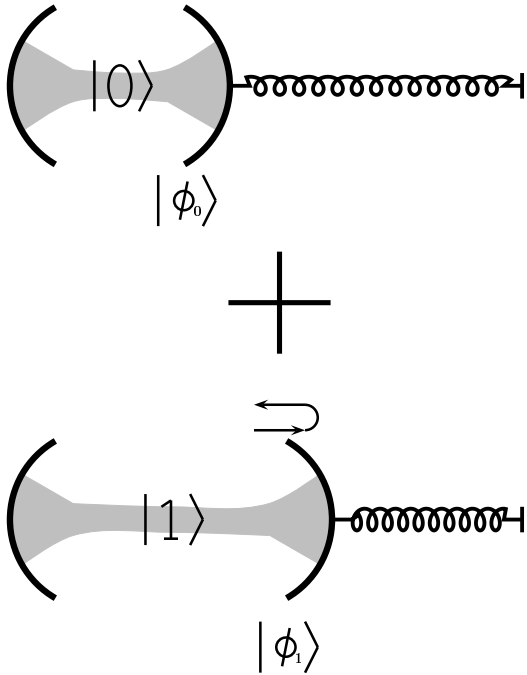}
\vspace{1cm}
\caption{\narrowtext The proposed opto-mechanical version
of Schr\"{o}dinger's thought experiment: the quantized single mode field is modified by the harmonic motion of one of the mirrors. If the mirror
started in a single coherent state then the result is an entangled state of the mirror motion and the cavity field, here labeled by $|0\rangle |\phi_0\rangle$ and $|1\rangle |\phi_1\rangle$. Given that the initial state of the mirror is a thermal
state, only a mixture of several such opto-mechanical cats with different
mean positions is produced.}
\label{mircat}
\end{center} 
\end{figure}

     The situation described by Eq.(\ref{cat}) between times $t=0$ and $t=2\pi/\omega_m$ is a mixture of several Schr\"{o}dinger's cat states of the type depicted in Fig.\ref{mircat} (where the
value of $n$ has been taken to be equal to $1$).
 Eq.(\ref{phi1}) implies that at $t=2\pi/\omega_m$, all states $|\phi_n (t)\rangle_{\mbox{\scriptsize m}}$ will
evolve back to $|\beta\rangle_{\mbox{\scriptsize m}}$ irrespective of $n$. Thus the mirror will return to its original thermal state (given by Eq.(\ref{th})) and the state of the cavity field will be disentangled from the mirror. In the absence of any EID, this state will be
\begin{equation}  
     |\psi(2\pi/\omega_m) \rangle_{\mbox{\scriptsize c}} = \frac{1}{\sqrt{2}}(|0\rangle_{\mbox{\scriptsize c}} - e^{i \kappa^2 n^2 2\pi}|n\rangle_{\mbox{\scriptsize c}})  .
\end{equation}

     In reality, two sources of decoherence will be present. The first one
is the decoherence due to photons leaking from the cavity and the second one is EID of the motional state of the mirror. The aim of this paper is to
show how the rate of the second type of decoherence can be determined. To
simplify our analysis, we shall take $n=1$ (i.e, the initial 
state inside the cavity is $\frac{1}{\sqrt{2}}(|0\rangle_{\mbox{\scriptsize c}} - |1\rangle_{\mbox{\scriptsize c}})$). First consider the case when no photon happens to leak out of the cavity up to a time $t$. If the damping constant of the cavity mirror is $\gamma_a$ then the probability
for this to happen is $\frac{1}{2}(1+e^{-\gamma_a t})$ \cite{barry}. In this case, the amplitude of the state $|1\rangle_{\mbox{\scriptsize c}}|\phi_1 (t)\rangle_{\mbox{\scriptsize m}}$ is suppressed with respect to the state $|0\rangle_{\mbox{\scriptsize c}}|\phi_0 (t)\rangle_{\mbox{\scriptsize m}}$ by a factor $e^{-\gamma_a t/2}$. In addition to this form of
decoherence, there is EID of the mirror's motional state. This has already been studied quite extensively, the basic result
being a rapid decay of those terms in the density matrix that are off-diagonal
in the basis of Gaussian coherent states \cite{zk,zk1}. However, the
diagonal terms in the coherent state basis are hardly affected on the
same time-scale (in fact, it has been
shown that in the case of a harmonic oscillator, coherent states emerge
as the most stable states under decoherence \cite{zk2}). Quite independent of
any specific model of decoherence, the satisfactory emergence of
classicality would require the off
diagonal terms in a coherent state basis to die much faster than the
diagonal terms as coherent states are the best candidates for classical
points in phase space. Thus, EID of the mirror's motional state 
 decreases the coherence between the states $|0\rangle_{\mbox{\scriptsize c}}|\phi_0 (t)\rangle_{\mbox{\scriptsize m}}$ and $|n\rangle_{\mbox{\scriptsize c}}|\phi_n (t)\rangle_{\mbox{\scriptsize m}}$ because $|\phi_0 (t)\rangle_{\mbox{\scriptsize m}}$ and $|\phi_n (t)\rangle_{\mbox{\scriptsize m}}$ are spatially separated coherent states of the mirror's motion. Let the average rate of this decoherence be $\Gamma_m$. Then the state of the system at time $t$ is given by

\begin{eqnarray}
\label{nol}
\rho(t)_{\mbox{\scriptsize {c+m}}} &=&(\frac{1}{1+e^{-\gamma_a t}}) \frac{1}{2 \pi \bar{n}} \int [\rho(t)_{00}-e^{-(\frac{\gamma_a}{2}+\Gamma_m)t}\rho(t)_{01} \nonumber \\
&-& e^{-(\frac{\gamma_a}{2}+\Gamma_m)t}\rho(t)_{10} \nonumber \\
&+& e^{-\gamma_a t}\rho(t)_{11}]_{\mbox{\scriptsize {c+m}}} \exp(-|\beta|^2/\bar{n}) ~ d^2\beta ,   
\end{eqnarray}
where the symbols $\rho(t)_{ij}$ denote states as given by Eqs.(\ref{e1})-(\ref{e4}). Eq.(\ref{nol}) shows that at time $t=2\pi/\omega_m$ the states of the cavity field and the mirror become dynamically disentangled. Now consider the complementary case (i.e when a photon
actually leaks out of the cavity between times $0$ and $t$). The total probability for this to happen is $\frac{1}{2}(1-e^{-\gamma_a t})$. As soon as the photon leaks out, the state of the cavity field goes to $(|0\rangle \langle 0|)_{\mbox{\scriptsize{c}}}$ and its state becomes completely disentangled from the
state of the mirror. Moreover, the mirror does not interact with the cavity
field any more as the interaction is proportional to the number operator of the
cavity field. Thus its state remains disentangled from the state of the cavity
field at all times after the photon leakage. Adding both the cases (photon loss
and no photon loss) with respective probabilities, one gets the state of the
cavity field at time $t=2\pi/\omega_m$ to be

\begin{eqnarray}
\label{cav}
\rho(2\pi/\omega_m)_{\mbox{\scriptsize {c}}} = \frac{(2-e^{-2\gamma_a \pi/\omega_m})}{2} |0\rangle \langle 0|_{\mbox{\scriptsize {c}}} &-& \nonumber \\ \frac{e^{-\gamma_a \pi/\omega_m}}{2} e^{-\Gamma_m 2\pi/\omega_m}(e^{i \kappa^2 2\pi}|1\rangle \langle 0|_{\mbox{\scriptsize {c}}} &+& e^{-i \kappa^2 2\pi}|0\rangle \langle 1|_{\mbox{\scriptsize {c}}})
\nonumber \\+ \frac{e^{-2\gamma_a \pi/\omega_m}}{2}|1\rangle \langle 1|_{\mbox{\scriptsize {c}}}. 
\end{eqnarray}
Thus by probing the state of the cavity field at time $t=2\pi/\omega_m$, it should be possible to determine the value
of $\Gamma_m$ and this gives the rate of decoherence of spatially separated
states of the macroscopic mirror.

           We should pause here to note that the state of the cavity field at $t=2\pi/\omega_m$ is completely {\em independent} of the initial thermal state (given by Eq.(\ref{th})), in which the mirror starts off. This feature is very important for our proposal. It implies that the effects on the cavity field
will be same {\em irrespective} of whether the mirror started off in a mixture
of coherent states (like a thermal state) or in a single coherent state. This makes the imprint of the demise of a single Schr\"{o}dinger's cat state on the cavity field identical to the imprint made by the demise of a mixture of several such states.

\vspace{4cm} 
\vspace{1cm} 

   The simplest method to determine the value of $\Gamma_m$ is to pass a
single two level atom (which interacts resonantly with the cavity field) in its ground state $|g\rangle$ through the cavity
at time $t=2\pi/\omega_m$, such that its flight time through the cavity is
half a Rabi oscillation period. The state of the cavity will get mapped  
onto the atom with $|e\rangle$ replacing $|1\rangle_{\mbox{\scriptsize {c}}}$ and $|g\rangle$ replacing $|0\rangle_{\mbox{\scriptsize {c}}}$ in Eq.(\ref{cav}). Then the probability of the atom to be in the
state $|+\rangle = \frac{1}{\sqrt{2}}(|g\rangle + |e\rangle)$ after it
exits the cavity, is 

\begin{equation}
\label{prob}
P(|+\rangle \langle+|) = \frac{1}{2}[1-e^{-\frac{\pi}{\omega_m}(\gamma_a+2\Gamma_m)} \cos{\kappa^2 2\pi}].
\end{equation}  
From the above equation it is clear that determining the probability of the exiting atom to be in the
state $|+\rangle$ will help us to determine the decoherence rate $\Gamma_m$ if
the order of magnitude of $\Gamma_m$ can be made greater than or about the same
as that of $\gamma_a$. Another requirement is that $\Gamma_m$ must be of the
same order as $\omega_m$ or even lower. Otherwise changes in $P(|+\rangle \langle+|)$ due to changes in $\Gamma_m$ would be too small to observe. Of
course, if one initially started with a general state $\frac{1}{\sqrt{2}}(|0\rangle_{\mbox{\scriptsize c}} - |n\rangle_{\mbox{\scriptsize c}})$ of the cavity field, then more general
tomography schemes \cite{wilk} will have to be used.

\section{A heuristic formula for the average decoherence rate}
   We now proceed to estimate $\Gamma_m$ in terms of the physical parameters of our system to illustrate the importance of this experiment from the point of view of testing the
decoherence of a macroscopic object. According to the models of references~\cite{zkpt} and~\cite{zk}, a superposition of coherent states spatially separated by a distance $\Delta x$ decoheres (when $\Delta x$ is greater than the thermal de Broglie wavelength $ \lambda_{th} = \hbar/\sqrt{2 m k_{\mbox{\tiny B}} \theta} $ ) on a
time-scale
\begin{equation}
\label{eqtd}
t_D = \frac{\hbar^2}{2 m \gamma_m k_{\mbox{\tiny B}} \theta (\Delta x)^2},
\end{equation}
where $m$ and $\gamma_m$ stand for the mass and damping constant of the object under consideration and $\theta$ is the temperature of the enclosure where
the object is placed.
 In our case, the spatial separation between the coherent states $|\phi_0 (t)\rangle_{\mbox{\scriptsize m}}$ and $|\phi_n (t)\rangle_{\mbox{\scriptsize m}}$ varies with time as

\begin{equation}
\label{del}
\Delta x (t) = \sqrt{\frac{\hbar}{2m\omega_m}} 2\kappa n (1-\cos{\omega_m t}) .
\end{equation}
Assuming, for the time being, that the decoherence process was entirely environment-induced, one can use Eqs.(\ref{eqtd}) and (\ref{del}) to calculate the average value
of the decoherence rate $\Gamma_m$ to be 
\begin{eqnarray}
\label{dec}
\Gamma_m &=& \frac{1}{(2\pi/\omega_m)} \frac{4\hbar \kappa^2 n^2}{2m\omega_m}\frac{2 m \gamma_m k_{\mbox{\scriptsize B}} \theta}{\hbar^2}\int_0^{2\pi/\omega_m} \!\!\!\!\!\!\!\!\!\!\!\!\!\!\! (1-\cos{\omega_m t})^2 dt 
\nonumber \\
         &=& \frac{3 n^2 \omega_0^2}{L^2 \omega_m^4 m} k_{\mbox{\scriptsize B}} \theta  \gamma_m ,
\end{eqnarray}
where the value of $\kappa$ has been substituted.

\section{Constraints on the parameters}
For our scheme to be successful in testing the decoherence of a macroscopic
object, and our method of analysis to be valid, certain parameter constraints have to be satisfied. The first constraint comes from Eq.(\ref{prob}). The decoherence rate to be measured, $\Gamma_m$,  has to be made
greater than or about the same order 
as that of the other decoherence rate $\gamma_a$. An associated requirement is that $\Gamma_m$ must be of the
same order as $\omega_m$ or even lower. This is because, in order to be able to measure a finite decoherence rate, we have to have only partial decoherence. If $\Gamma_m$ is much greater than $\omega_m$ then the decoherence will be
too fast and essentially complete before even one oscillation period of the mirror and thereby not measurable. Thus we have,

  {\bf Constraint $1$:}
\begin{equation} 
\omega_m \sim \Gamma_m \geq \gamma_a   \nonumber \\
\end{equation}

  The next constraint is required for our heuristic
treatment of the decoherence of the mirror to be valid. The use of Eq.(\ref{eqtd}) is valid only when $\Delta x$ is greater than the thermal de Broglie wavelength $ \lambda_{th} = \hbar/\sqrt{2 m k_{\mbox{\tiny B}} \theta} $. Using the expression for $ \lambda_{th} $ in Eq.(\ref{del}) we get

 {\bf Constraint $2$:}
\begin{equation} 
\frac{\omega_0^2}{L^2 \omega_m^4 m}k_{\mbox{\scriptsize B}} \theta >> 1 \nonumber \\
\end{equation}
 
A final constraint comes from the fact that the spatial separation  $\Delta x$
between the superposed peaks must be greater than or at least of the 
same order as the width of a single peak. This is
a requirement for two reasons: firstly for the validity of our heuristic treatment of decoherence, and secondly for the components of the Schr\"{o}dinger's cat to be sufficiently separated in space (i.e at least
as much separated than the spatial width of each component of the Schr\"{o}dinger's cat). As the width of each of the components of the cat
is simply equal to the width of a coherent state, using Eq.(\ref{del}) and the
fact that $n=1$, we get,

{\bf Constraint $3$:}
\begin{equation} 
\kappa \geq 1 \nonumber \\
\end{equation}

We should stress that while constraint $1$ will be a {\em necessary} constraint
in any analysis of our system, constraints $2$ and $3$ really arise due to
our method of analysis. If we could calculate the decoherence rate when the
superposed wavepackets almost overlap each other, then neither of the constraints $2$ or $3$ would be needed. But in that case, the decrease in the
decoherence rate may be so much that constraint $1$ becomes difficult to achieve. We leave the analysis of this domain for the future. We now proceed to
propose a set of parameters which satisfy the above constraints and which
are fairly close to those currently realisable.

\section{Experimental parameter regimes satisfying the constraints}
     
        At first, let us briefly state the available ranges of the various parameters
involved in our experiment as far as the technology stands today. The frequency
of mechanical oscillators ($\omega_m/2\pi$ in our case) is normally in the KHz domain, but can be
made to rise up to a GHz \cite{rk}. However, in the case of such high frequencies, the mass of the oscillators is very small (about $10^{-17}\;\mbox{kg}$ \cite{rk}). The mass $m$ of the movable mirror has
no upper restriction, but is bounded on the lower
side by the requirement of having to support the beam waist of an electromagnetic field mode. This means that the masses of mirrors for microwave cavities
should be no smaller than about $0.1 \mbox{g}$ while those for optical cavities can
go as low as $10^{-15} \mbox{kg}$. The length $L$ of the
cavity can be no lower than $1\;\mbox{cm}$ in the microwave domain but can be as low as
$1\; \mu\mbox{m}$
in 
the optical domain. In fact, optical cavities 
with a length
of the 
order of 
$10 \; \mu\mbox{m}$ 
already 
exist \cite{kim}. While there is 
essentially no limit to how high the mechanical damping rate, $\gamma_m$, of the moving mirror can
be made, there is a lower limit (not necessarily a fundamental limit,
but the best achievable in current experiments). Oscillating cavity mirrors
with $\omega_m/2\pi \sim 10~\mbox{kHz}$ and Q-factor $\sim 10^{6}$ have been
fabricated \cite{schil}. We will take the corresponding $\gamma_m \sim 10^{-2} \mbox{s}^{-1}$ to be a lower limit on the value of the mechanical damping constant. The lowest temperature $\theta$ to which a macroscopic mirror has been
cooled as yet, is about $0.5 \mbox{K}$ \cite{Rmp}. As far as the damping
constant $\gamma_a$ due to leakage of photons from the cavity is concerned,
the lowest values are $10^{7}~\mbox{s}^{-1}$ for optical cavities of $L \sim 10 \; \mu\mbox{m}$ (with stationary mirrors) \cite{kim},  $10^{6}~\mbox{s}^{-1}$ for optical cavities of $L \sim 1 \; \mbox{cm}$ (with a moving mirror) \cite{schil}, and $10 \mbox{s}^{-1}$ for microwave cavities of $L \sim 1 \; \mbox{cm}$ (with a stationary mirror) \cite{Rmp}.

                   Now let us examine a parameter regime in which all our constraints are satisfied. We first look at optical cavities ($\omega_0/2\pi \sim 10^{15} \mbox{Hz}$). For optical cavities we can choose $L \sim 10 \; \mu\mbox{m}$ \cite{kim}. We choose $m=1 \;\mbox{mg}, \gamma_m=10^{-2} \;\mbox{s}^{-1}, \omega_m/2\pi=10 \; \mbox{kHz}$, and $\theta=0.1 \; \mbox{K}$. With this choice of parameters, 

\begin{equation} 
\frac{\omega_0^2}{L^2 \omega_m^4 m}k_{\mbox{\scriptsize B}} \theta \sim 10^{6} \nonumber \\
\end{equation}

and 

\begin{equation} 
\kappa \sim 1 \nonumber .\\
\end{equation}
So both the constraints 2 and 3 are satisfied. Also we have

\begin{equation} 
\Gamma_m \sim \omega_m \sim 10^4 \;\mbox{s}^{-1} .
\end{equation}

Thus, in order to satisfy constraint 3, we require $\gamma_a \leq 10^4 \;\mbox{s}^{-1}$. While this value of $\gamma_a$ is only three orders of magnitude removed from the
best reflectivity for stationary mirrors and five orders of magnitude
removed from the best reflectivity for moving mirrors in the optical domain at present,
all the other parameters assumed here are well within the experimentally
accessible domain. We don't see any {\em fundamental} reason why the reflectivity of the moving mirror cannot be increased by a few orders of
magnitude, as the mirror is quite macroscopic (of milligram mass). Note that in the above
case the position separation $\Delta x$ between the superposed components
is {\em really} tiny (of the order of $10^{-16} \mbox{m}$ !), yet even this is 
sufficient to cause an observable rate of decoherence. This is because the
macroscopic nature of the moving mirror implies that even this minute separation
is much larger than the thermal de Broglie wavelength. We note that one is allowed
to increase the mass of the moving mirror to about $100~\mbox{mg}$, if the
length of the cavity is decreased to $1~\mu\mbox{m}$. Our constraints will still have exactly the same values as above when this change is made. However, it seems that $100~\mbox{mg}$ is probably the largest mass the mirror in our
experiment can possibly have. Note that mirror 
masses of $1~\mbox{mg}-100~\mbox{mg}$ are well within experimentally
accessible domains as mirror masses of the order of $10~\mbox{mg}$
have already been used in optical bistability experiments \cite{Meyes}.

                     The above choice of parameters was entirely motivated
by an attempt to keep the parameters as close as possible to 
an existing optical cavity with a moving mirror experiment \cite{schil}. Our constraints require the mirror reflectivity of this experiment to be improved in order for our
proposal to be a success. However, another alternative is to keep the
values of mirror reflectivity same as in existing experiments but move on to
a mirror oscillating at a much higher frequency. Let us choose $\omega_m \sim
\gamma_a \sim 10^7 \;\mbox{s}^{-1}$ (though this value of $\gamma_a$ is for the
best existing static mirror). To make $\Gamma_m \sim 10^7$ and satisfy constraint 1, we require to choose low $L \sim 10 \; \mu\mbox{m}$, low $m \sim
10^{-15} \mbox{kg}$, temperature
$\theta \sim 10 \mbox{K}$ and high $\gamma_m \sim 100 \mbox{s}^{-1}$. The frequency of the
cavity mode is kept the same ($\omega_0/2\pi \sim 10^{15}~\mbox{Hz}$). This choice
also satisfies constraints 2 and 3 as

\begin{equation} 
\frac{\omega_0^2}{L^2 \omega_m^4 m}k_{\mbox{\scriptsize B}} \theta \sim 10^{5} \nonumber \\
\end{equation}

and 

\begin{equation} 
\kappa \sim 1 \nonumber .\\
\end{equation}

Among the basic changes made here from existing experiments, the temperature
and higher $\gamma_m$ will only be too easy to achieve. However, a cavity mirror with a very tiny
mass of $10^{-15} \mbox{kg}$ should be difficult to fabricate. But
mechanical resonators of much lower mass have already been
fabricated \cite{rk}. Moreover, there is
nothing of {\em principle} which excludes such a mirror for an optical cavity
because it can still support an optical beam waist. Besides, small masses are
also required for mechanical resonators of very high frequencies \cite{rk} as
in this case. One might also think that the very small time period of mirror
oscillation ($10^{-7} \mbox{s}$) may be a barrier to the tomography of the
cavity field using atoms. But cesium atoms with lifetime $\sim 10 \mbox{ns}$
should be useful for this purpose.

   We should now proceed to examine the prospects of implementing our
experiment successfully in the microwave domain. In this domain the
lowest possible values of $L$ and $m$ are already fixed to be $1 \mbox{cm}$
and $0.1 \mbox{g}$. Thus constraint 3 implies that the maximum
value $\omega_m/2\pi$ can take is $10^{-2} ~\mbox{Hz}$. But this clearly makes
constraints 1 and 2 impossible to satisfy unless

\begin{equation}
 \theta  \gamma_m < 10^{-14} \mbox{K s}^{-1}.
\end{equation}
This value of the $\theta  \gamma_m$ product lies well outside potentially
realizable domains. Thus strictly speaking, an experiment of the kind proposed
here would not achieve success in the microwave domain. The only way it could,
would be to use a much smaller mass for the oscillating mirror. As such a mirror
will not be able to support a microwave cavity mode, we would have to introduce
it as a small mechanical resonator inside a cavity with fixed mirrors. This 
should be quite an interesting but different problem to study, because the
cavity field--mechanical resonator interaction Hamiltonian in this case may be 
different. 

\section{Parameter regime required to test gravitationally induced
collapse theories}

   We may describe the parameters used in the estimations so far as being potentially realizable. Let us now identify the range of parameters that would be required if one intends to extend the scope of our experiment to test the gravitationally induced
objective reduction (OR) models of the type proposed by Penrose and Diosi~\cite{pen}.
 According to this model the decoherence rate will be 
\begin{equation}
\gamma_{OR} \sim \frac{E}{\hbar}
\end{equation}                 
where $E$ is the mean field gravitational interaction energy. We will examine only the case in which $\Delta x < R$, where $R$ is the dimension of the object, as
this is the easiest to achieve experimentally. In the case of a spherical
geometry of the mirror (we use such an assumption just for an estimate)
$ E \sim G m^2 (\Delta x)^2 / R^3$. Using the expression for $\Delta x$ from Eq.(\ref{del}) and substituting $R^3$ by $m/D$, where $D$ is the density of the object, one gets,
\begin{equation}
\label{or}
\gamma_{OR} \sim \frac{n^2 \omega_0^2}{L^2 \omega_m^4 m} G \hbar D  .
\end{equation}
Comparison of Eqs.(\ref{dec}) and (\ref{or}) shows that decoherence rates according
to EID and OR have exactly the same dependence on parameters $L, m, \omega_m, \omega_0$ and $n$ and therefore one cannot distinguish between these models by varying any of these parameters (Of course this statement is true only for
a spherical geometry of the mirror). In order to reduce the effect of EID to such an
extent that effects of OR become prominent one needs
\begin{equation}
G \hbar D > k_{\mbox{\scriptsize B}} \theta  \gamma_m .
\end{equation} 
Taking the density $D$ of a typical solid to be of the order of $10^3$ kg $\mbox{m}^{-3}$, one gets 
\begin{equation}
 \theta  \gamma_m < 10^{-19} \mbox{K s}^{-1}.
\end{equation}
Currently, temperature of a macroscopic object can be brought down to at most
$0.1$ K and a fairly optimistic estimate of $\gamma_m$ is $10^{-2} \mbox{s}^{-1}$ (a mechanical oscillator that dissipates its energy in about $100$ s). Thus an improvement of the product $\theta\gamma_m $ by sixteen orders of magnitude would be necessary to test OR using our scheme.

\section{Conclusions}
In conclusion, we note that the experiment we have proposed just applies Schr\"{o}dinger's method to a realistic system of a cavity field and a
macroscopic moving mirror. Of course, to achieve Schr\"{o}dinger's original
aim (creating a macroscopic superposition), our scheme will have to be
combined with a scheme that prepares the mirror in a pure coherent state. 
However for testing the rules of decoherence of a macroscopic object, our scheme
is sufficient. A special feature of our scheme is that the microscopic system (i.e the cavity field) which creates the mixture of Schr\"{o}dinger's cats is itself being used later as a kind of meter to read the decoherence that the mirror undergoes while the two systems are entangled. We believe that this is a canonical
system for systematic probing of decoherence and offers an extensive scope
of further work from both theoretical and experimental points
of view. Modelling the EID of our system starting from the very first principles (assuming different types of coupling and environment) is necessary to check the
accuracy of formula (\ref{dec}). A variant of our set up in which a small mechanical resonator is introduced inside a cavity should be an 
interesting problem to study. There can be an entire range of masses for such a mechanical oscillator introduced inside a cavity: starting from trapped ions \cite{mon}, to trapped molecules and nanoparticles \cite{wei}, to the smallest mechanical resonators that can be fabricated \cite{rk}. There can also be other variants of our proposal such as extending schemes in which an 
atom trapped in a cavity produces Fock states \cite{lw} to 
include the effects of a moving mirror. 

    The experimental challenge is in either of the two
directions: to improve the reflectivity of existing 
macroscopic mirrors or to decrease the mass of the mirrors 
without decreasing the existing reflectivity. There is nothing
of {\em principle} which prohibits increasing the reflectivity of a
macroscopic mirror, nor any {\em fundamental} connection between mirror
reflectivity and mass (as long as the mirror can support the beam waist).
So we don't see any real obstacle in progress directed at the
possible realization of our proposal.

    We would like to end with a note clarifying the exact relevance
of our experiment. It is much more than detecting the presence of
a thermal environment around the system. We are really interested
in detecting {\em how} this environment causes the demise of the coherence
between superposed spatially separated states of a macroscopic object. This is
interesting, because {\em irrespective} of any role it plays
in the foundations of quantum mechanics, thermal environment induced
decoherence is a {\em real} phenomenon yet to be systematically
probed in the macroscopic domain. As far as the relevance of mentioning
OR in this paper is concerned, it is mainly to emphasize the degree of 
technological improvement necessary in order to bring such effects into
the observable domain. This technological feat should be taken up as a
challenge unless shown to be impossible by some fundamental principle
of physics.

\section{Acknowledgements}
This work was supported in part by the UK Engineering and Physical Sciences Research Council, the European Union, the British Council, the Inlaks Foundation and the New Zealand Vice Chancellor's Committee. One of us (SB) would like to acknowledge a very long interaction with Dipankar Home on foundational issues of quantum mechanics. We acknowledge conversations with 
M. Blencowe, B. L. Hu, M. S. Kim, G. Kurizki, R. Penrose, I. C. Percival, W. Power, S. Schiller and  W. H. Zurek.

\end{multicols}

\newpage

\end{document}